\documentclass[numberedappendix,twocolappendix,iop,apj]{emulateapj}

\usepackage{graphics}
\usepackage{epstopdf}
\usepackage{footnote}
\usepackage{tablefootnote}
\usepackage{lipsum}
\usepackage{float}
\usepackage{bm}
\makesavenoteenv{tabular}
\usepackage{psfig,amsfonts,amsmath,graphicx,natbib,url,hyperref}
\usepackage{amssymb}
\usepackage{txfonts}

\usepackage{booktabs}
\hypersetup{colorlinks=true, urlcolor=blue, citecolor=blue}

\newcommand{\nua}{\ensuremath{\nu_{\rm a}}}

\newcommand{\numax}{\ensuremath{\nu_{\rm m}}}

\newcommand{\nuc}{\ensuremath{\nu_{\rm c}}}
\newcommand{\nuaNT}{\ensuremath{\nu_{\rm a, NT}}}

\newcommand{\numaxNT}{\ensuremath{\nu_{\rm m, NT}}}

\newcommand{\nuaT}{\ensuremath{\nu_{\rm a, T}}}

\newcommand{\numaxT}{\ensuremath{\nu_{\rm m, T}}}

\newcommand{\fT}{\ensuremath{f_{\rm T}}}
\newcommand{\fNT}{\ensuremath{f_{\rm NT}}}
\newcommand{\fmax}{\ensuremath{f_{\rm max}}}

\shorttitle{Thermal Electrons}
\shortauthors{Ressler \& Laskar}

\def\ucb{1}
\def\nrao{2}

\begin{document}

\title{Thermal Electrons in GRB Afterglows}

\author{Sean M. Ressler\altaffilmark{\ucb} and Tanmoy Laskar\altaffilmark{\nrao,\ucb}}
\altaffiltext{\ucb}{Department of Astronomy, University of California, 501 Campbell Hall, 
Berkeley, CA 94720-3411, USA} 
\altaffiltext{\nrao}{National Radio Astronomy Observatory, 520 Edgemont Road, Charlottesville, VA 
22903, USA} 


\begin{abstract}
To date, nearly all multi-wavelength modeling of long-duration $\gamma$-ray bursts has ignored synchrotron radiation from the significant population of electrons expected to pass the shock without acceleration into a power-law distribution. We investigate the effect of including the contribution of thermal, non-accelerated electrons to synchrotron absorption and emission in the standard afterglow model, and show that these thermal electrons provide an additional source of opacity to synchrotron self-absorption, and yield an additional emission component at higher energies. The extra opacity results in an increase in the synchrotron self-absorption frequency by factors of 10--$100$ for fiducial parameters. The nature of the additional emission depends on the details of the thermal population, but is generally observed to yield a spectral peak in the optical brighter than radiation from the nonthermal population by similar factors a few seconds after the burst, remaining detectable at millimeter and radio frequencies several days later.
\end{abstract}

\maketitle

\section{Introduction}
Observations of synchrotron radiation in supernovae, $\gamma$-ray bursts (GRBs), and tidal disruption events provide an efficient means to probe high-energy transient phenomena. Multi-wavelength observations of these events have been successfully leveraged to determine the energy, particle density, and magnetic field strength, thereby constraining the progenitors and physical processes powering these extreme explosions \citep{spn98,pk02,sckf06,mgm12}.

The observed radiation in these sources arises from electrons accelerated to relativistic energies in shocks. While the method of accelerating these electrons is far from understood, all proposed models predict that only a fraction $\fNT < 1$ of electrons are accelerated into the non-thermal distribution responsible for the observed radiation, the rest forming a relativistic Maxwellian distribution at lower energies. The precise value of \fNT\ depends strongly on the shock magnetization, and is expected to range from $\lesssim1\%$ to $\approx15\%$ for different shock heating mechanisms \citep{ss11}. A determination of \fNT\  would directly constrain the properties of the shocks generating synchrotron radiation in astrophysical sources, a vital step toward understanding this ubiquitous radiation process. 

Unfortunately, \fNT\ is completely degenerate with the other physical parameters governing the observed spectra and light curves from astrophysical transients. Previous studies 
have therefore simply
employed a value of unity. This degeneracy has also caused an unknown degree of uncertainty in the values of the physical parameters derived from fitting observations of these sources.
\cite{Eichler2005} first clarified this degeneracy in the context of GRB afterglows. Their pioneering work, which included a rudimentary treatment of synchrotron cooling and self-absorption, suggested the thermal population could be distinguished by an early `pre-brightening' at radio wavelengths. Both \cite{gs09} and \cite{webn17} studied the effect of thermal electrons on the observed spectrum, the latter including the effects of pion decay and inverse Compton radiation.  Both, however, ignored synchrotron self-absorption, which is expected to affect radio observations. \cite{Toma2008} further studied the radio polarimetry signature of a mono-energetic thermal peak in the distribution function, finding that thermal electrons may suppress Faraday rotation if the magnetic field is well ordered.

In this work, we study the effect of non-shock heated electrons on the synchrotron radiation from GRB afterglows, including synchrotron cooling and self-absorption.
To parameterize the uncertainty of the thermal heating by shocks in collisionless plasmas, we consider two extreme cases: 1) strong shock heating, where the post-shock electron temperature is comparable to the post-shock gas temperature, and 2) weak shock heating, where the post-shock electron temperature is much less than the post-shock gas temperature. We study the effect of inefficient acceleration as a function of the accelerated fraction \fNT, focusing on a fiducial set of parameters typical of observed GRB afterglows. 



\section{Physical Model}
Our underlying physical model is the self-similar hydrodynamic solution for a spherically symmetric point explosion \citep{BM1979}. We use the framework of \cite{GS2002} (hereafter, GS02) to describe synchrotron radiation from the resulting relativistic shock. We modify the distribution function of electrons downstream of the shock to include contributions from thermal (non-accelerated) particles. The parameters in our model are the explosion energy, $E$, the external number density, $n_{ext}$, the fraction of internal energy given to nonthermal electrons, $\epsilon_e$, the fraction of internal energy given to the magnetic field, $\epsilon_B$, the nonthermal power law index, $p$, and \fNT. 
We describe our modifications to the distribution function in Section \ref{edist}; for a more detailed description of the hydrodynamics, we refer the reader to GS02.

\subsection{Fluid Solution}
The \citet{BM1979} solution depends on the initial explosion energy, $E$, and the ambient medium density, $n_{ext}$. In terms of these parameters, the Lorentz factor of the shock can be expressed as a power law in the coordinate time, $t$ as
\begin{equation}
\Gamma = \sqrt{\frac{17 E}{8 \pi n_{ext} m_p c^5 t^3}},
\end{equation}
corresponding to a shock radius, 
\begin{equation}
R = c t \left(1 + \frac{1}{8\Gamma^2} \right),
\end{equation}
where $m_p$ is the proton mass and $c$ is the speed of light. 
Note that here and throughout we keep only the leading order terms in $\Gamma$.  The post-shock fluid is characterized by the similarity variable, $\chi$, which is related to the coordinate time, $t$, coordinate radius, $r$, and the shock radius:
\begin{equation}
\chi = 1 + 8 \Gamma^2 \left(\frac{R-r}{R}\right).
\label{eq:chi_def}
\end{equation}
In terms of this variable, the rest frame internal energy, $e$, Lorentz factor, $\gamma$, and rest frame number density, $n$, take the simple form 
\begin{eqnarray}
\begin{aligned}
e &= 2 \Gamma^2 m_p c^2 n_{ext}  \chi^{-17/12} \\
\gamma &= 2^{-1/2} \Gamma \chi^{-1/2} \\
n &= 2^{3/2} \Gamma n_{ext} \chi^{-5/4}.
\end{aligned}
\end{eqnarray}
We assume the post-shock magnetic field is isotropic, and carries a fixed fraction, $\epsilon_B$ of the shock energy,
\begin{equation}
B = \sqrt{ 8 \pi \epsilon_B e},
\end{equation}
Finally, in order to evolve the Lorentz factor of an individual electron as it travels away from the shock (see \S \ref{sec:Nevol}), we need not only the solution for each fluid element at a fixed coordinate time and radius, but also the fluid properties that it possessed at the time $t_0$ when it initially crossed the shock. We label these quantities with a subscript `0.' These can be derived by solving for the worldline of the fluid element using the relation
$\partial r/\partial t = \beta c \approx   c( 1 - \gamma^{-2} / 2)$, which yields (GS02):
\begin{eqnarray}
\begin{aligned}
e_0 &= e \chi^{13/6}  \\
\gamma_0 &= \gamma \chi^{7/8} \\
n_0 &= n  \chi^{13/8} \\
B_0 &= B \chi^{13/12}.
\end{aligned}
\end{eqnarray}

\begin{figure*}
\centering
\includegraphics[width=0.45\textwidth]{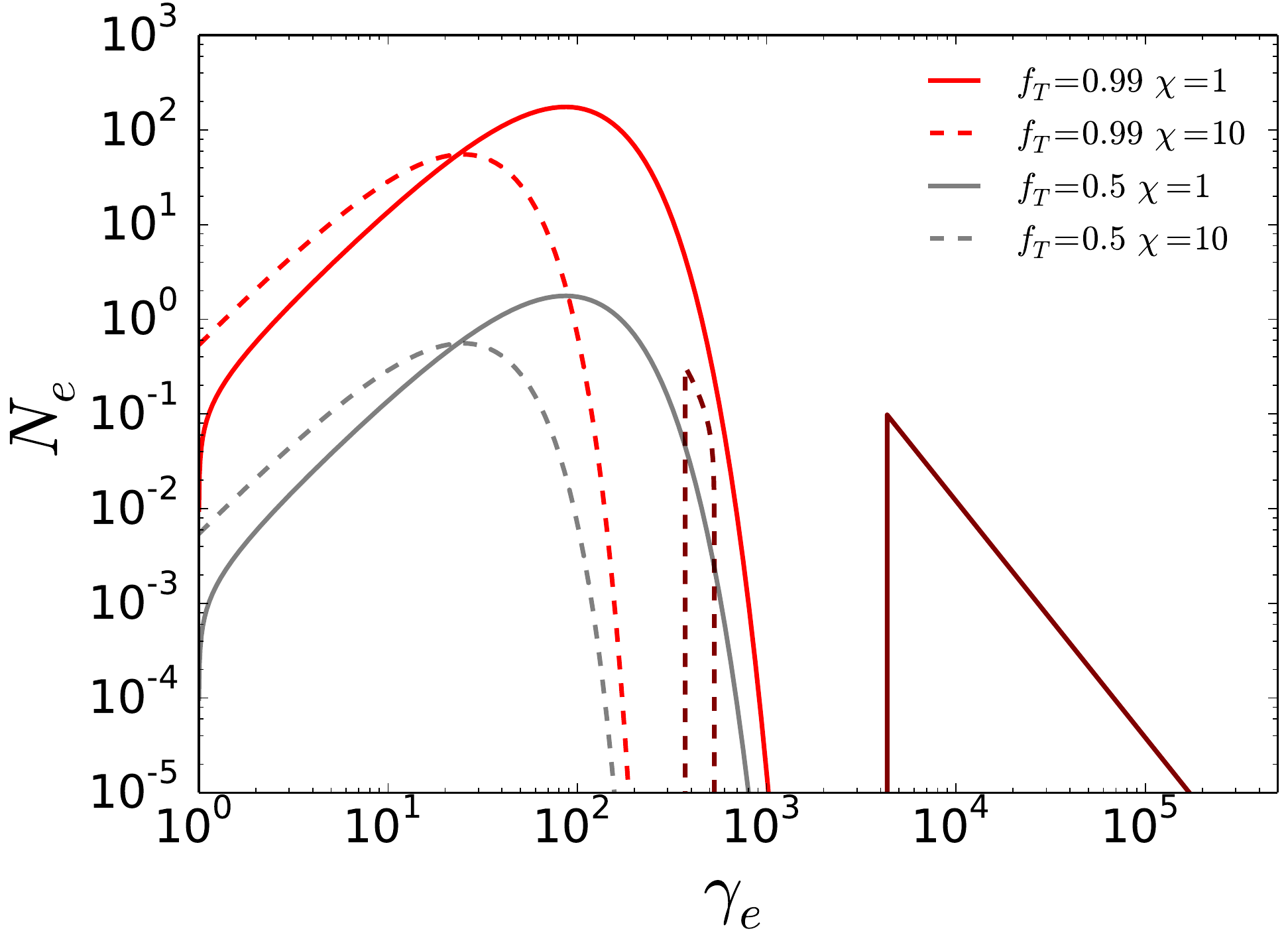}
\includegraphics[width=0.45\textwidth]{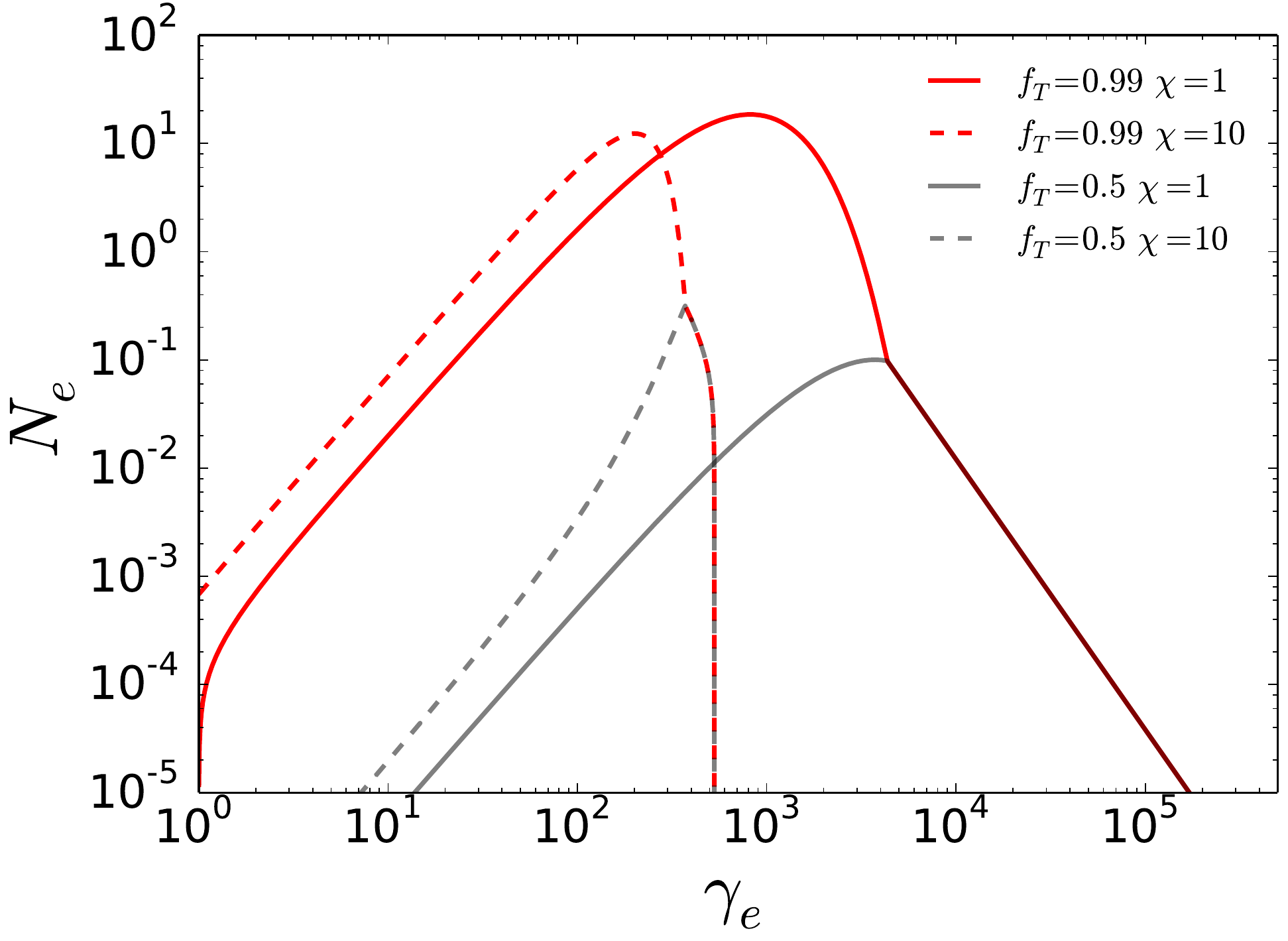}
\caption{Representative examples of the electron distribution functions for the `cold electron model' (left) and the `hot electron model' (right) for different choices of the fraction of thermal electrons, \fT, and at two different times along the world line of a fluid element.  $\chi = 1$ represents the distribution just behind the shock at a time $t_0 \approx 30 $ days, while $\chi = 10$ corresponds to a later time, $t \approx 53$ days, when the fluid element is further away from the shock and has been subject to radiative and adiabatic losses.
The fluid parameters are chosen to scale with \fNT\ such that the nonthermal distribution is independent of \fNT, thus the nonthermal parts of the distribution for different \fT\ are coincident.  
}
\label{fig:dist}
\end{figure*}

\subsection{Electron Distribution Function}
\label{edist}
We consider two classes of phenomenological models for the electron distribution function, 
in both of which
a fraction \fNT\ of the electrons are accelerated from a thermal distribution to a power law at the shock front. The remaining fraction, $\fT = 1-\fNT$, are either efficiently shock-heated (the ``Hot Electron Model'') or inefficiently shock-heated (the ``Cold Electron Model'').  As the electrons advect away from the shock front, the distribution suffers radiative and adiabatic losses.
We ignore the negligible effect of Coulomb collisions. The post-shock electron distribution can be split into a thermal and nonthermal components, written as a function of the electron Lorentz factor, $\gamma_e$, and characterized by a power law index, $p$, a minimum accelerated electron Lorentz factor, $\gamma_{\rm min,0}$, and normalization constants $K_{0,th}$ and $K_{0,nt}$:
\begin{equation}
 N_{e,th}  = K_{0,th} \frac{\gamma_e^2 \beta }{\Theta_{e,0} K_{2}\left(1/\Theta_{e,0}\right)} e^{-\gamma_e/\Theta_{e,0}} \approx K_{0,th} \frac{\gamma_e^2}{2\Theta_{e,0}^3} e^{-\gamma_e/\Theta_{e,0}}
\label{eq:thermal_0}
\end{equation}
 for $\gamma_e<\gamma_{\rm min,0}$,  and 
\begin{equation}
N_{e,nt} = K_{0,nt} \gamma_e^{-p}
\label{eq:NT_0}
\end{equation}
for $\gamma_e>\gamma_{\rm min,0}$, with 
\begin{equation}
\gamma_{\rm min,0} = \frac{p-2}{p-1}\frac{\epsilon_e e_0 }{ \fNT n_0 m_e c^2}.
\label{eq:gamma_min}
\end{equation}
Here $m_e$ is the electron mass, $\beta \equiv \sqrt{1-1/\gamma_e^2}$, $K_2(1/\Theta_{e,0})$ is a modified Bessel function of the 2nd kind. For the relativistic electrons considered here, $\Theta_{e,0} \gg 1$, $\beta \approx 1$, and $K_2(1/\Theta_{e,0}) \approx 2 \Theta_{e,0}^2$. The difference between the two models lies in the prescriptions for $\Theta_{e,0}$, which we now describe.

\subsubsection{Cold Electron Model}
\label{sec:cold}

In the cold electron model, inefficient shock heating of thermal electrons results in a post-shock electron temperature that is much less than the gas temperature.
These thermal electrons have Lorentz factors, $\gamma_e m_e\ll \gamma m_p$.  The nonthermal particles, on the other hand, are considered to be accelerated to a minimum Lorentz factor of $\gamma_{\rm min,0} m_e \sim \gamma m_p$, so that the two components of the distribution are clearly separated. We parameterize $\Theta_{e,0}$ with the fractional temperature of the thermal particles, $\eta_e \equiv T_e/T_g$:
\begin{equation}
\Theta_{e,0} = \frac{\eta_e}{\gamma_{\rm ad}} \frac{e_0}{n_0 k_b m_e c^2},
\label{eq:theta_eta}
\end{equation}
where $T_g$ is the gas temperature, $k_b$ is the Boltzmann constant, and $\gamma_{\rm ad}=4/3$ is the gas adiabatic index. To normalize the distribution functions, we impose:
\begin{equation}
\int\limits_{0}^{\infty} N_{e,th} d\gamma_e = \fT n_0
\label{eq:th_norm}
\end{equation} 
and 
\begin{equation}
\int\limits_{\gamma_{\rm min,0}}^{\infty} N_{e,nt} d\gamma_e = \fNT n_0,
\label{eq:nt_norm}
\end{equation}
yielding $K_{0,th} = \fT n_0$ and $K_{0,nt} = \fNT n_0 (p-1) \gamma_{\rm min,0}^{p-1}$. 

For larger values of $\eta_e$,  the nonthermal and thermal distributions overlap, causing a discontinuity in the electron distribution at $\gamma_{\rm min,0}$. To avoid this unphysical feature, we limit the use of this model to $\eta_e < \eta_{max}$, where $\eta_{max}$ (corresponding to some $\Theta_{e,0,max}$) is determined by requiring that 99\% of the thermal electrons have Lorentz factors below $\gamma_{\rm min,0}$, so that
\begin{equation}
\int\limits_0^{\gamma_{\rm min,0}} N_{e,th}(\gamma_e,\Theta_{e,0,max}) = 0.99 \fT n_0,
\label{eq:th_max}
\end{equation}
giving
\begin{equation}
\eta_{max} \approx 0.1 \left[3 \frac{p-2}{p-1} \frac{\epsilon_e}{\fNT}\right] \Leftrightarrow \Theta_{e,0,max} \approx 0.1 \gamma_{min,0}.
\label{eq:eta_max}
\end{equation}

\subsubsection{Hot Electron Model}
\label{sec:hot}

The hot electron model is characterized by efficient heating of thermal electrons at the shock, so that the post shock electron temperature is of order the total gas temperature, with typical Lorentz factors of $\gamma_e m_e \sim \gamma m_p$.  Since the nonthermal particles are also assumed to have Lorentz factors above $\gamma_{\rm min,0}m_e \sim \gamma m_p$, there is significant overlap between the two components of the distribution function. To connect the two, we adopt the prescription of \citet{Yuan2003} (similarly, \citealt{gs09}). 
We require 
$N_{e,th}(\gamma_{\rm min,0}) = N_{e,nt}(\gamma_{\rm min,0})$
which, when combined with the normalization conditions
\begin{equation}
\int\limits_{0}^{\gamma_{\rm min,0}} N_{e,th} d\gamma_e = \fT n_0
\end{equation} 
and 
\begin{equation}
\int\limits_{\gamma_{\rm min,0}}^{\infty} N_{e,nt} d\gamma_e = \fNT n_0,
\end{equation}
provides a set of nonlinear equations for $y_{\rm min} \equiv \gamma_{\rm min,0}/\Theta_{e,0}$ and $K_{0,th}$, while $K_{0,nt} = \fNT n_0 (p-1) \gamma_{\rm min,0}^{p-1} $. $y_{\rm min} $ is then determined by solving the  transcendental equation:
\begin{equation}
\frac{y_{\rm min}^3 e^{-y_{\rm min}}}{2 - e^{-y_{\rm min}}\left(y_{\rm min}^2 + 2y_{\rm min} + 2\right)}= \frac{\fNT}{\fT}\left(p-1\right),
\label{eq:ymin}
\end{equation}
which is then directly related to  the normalization constant for the thermal distribution:
\begin{equation}
K_{0,th} =  \frac{2\fT n_0 }{2 - e^{-y_{\rm min}}\left(y_{\rm min}^2 + 2y_{\rm min} + 2\right)}.
\end{equation}
For a fixed $\gamma_{min,0}$, as \fT\ gets smaller (\fNT\ gets larger), the value of $\Theta_{e,0}$ needed to continuously match the distribution functions gets larger.  As $\Theta_{e,0} \rightarrow \infty$, the thermal electron distribution approaches a  $\gamma_e^2$ power law distribution with only one free parameter, making the system over-determined.  This happens at a finite value of \fNT, denoted \fmax, which is given by the solution to equation \eqref{eq:ymin} as $y_{\rm min}\rightarrow 0$:
\begin{equation}
\fmax = \frac{3}{p+2}.
\label{eq:fmax}
\end{equation}
For $\fNT>\fmax$, the thermal distribution function cannot be continuously connected to the nonthermal distribution function while still maintaining the proper normalization.

Finally, we note that in this model $\eta_e$ as defined in \eqref{eq:theta_eta} is no longer a free parameter but directly related to $\epsilon_e$:
\begin{equation}
\eta_e = 3  \frac{p-2}{p-1}\frac{\epsilon_e}{\fNT y_{\rm min}}.
\end{equation}

\subsubsection{Evolution of the Electron Distribution Function}
\label{sec:Nevol}
The above prescriptions are applied to the electron distribution just behind the shock.  As the fluid advects away from the shock, each individual electron is subject to radiative and adiabatic losses,
\begin{equation}
\frac{d \gamma_e}{d t'} = - \frac{\sigma_{\rm T}B^2}{6 \pi m_e c} \gamma_e^2 + \frac{\gamma_e}{3n} \frac{dn}{dt'},
\label{eq:dgamdt}
\end{equation}
where $t'$ is measured in the frame of the post-shock fluid. The solution to this equation gives the electron Lorentz factor as a function of its Lorentz factor at the shock, $\gamma_{e,0}$, and the similarity variable, $\chi$: $\gamma_e(\gamma_{e,0},\chi)$. As the electrons lose energy, the distribution function will evolve 
conserving particle number:
\begin{equation}
\frac{N_e(\gamma_e)}{n} d\gamma_e = \frac{N_e(\gamma_{e,0})}{n_0} d\gamma_{e,0},
\end{equation}
so that the distribution function at an arbitrary point in the flow can be written as 
\begin{equation}
N_e(\gamma_e) = \frac{n}{n_0} N_e(\gamma_{e,0}) \frac{d\gamma_{e,0}}{d\gamma_e}.
\label{eq:Ne_evol}
\end{equation}
We use Equation \eqref{eq:Ne_evol} to evolve both the nonthermal and the thermal distribution functions.  The minimum Lorentz factor for the nonthermal particles is evolved according to 
equation \eqref{eq:dgamdt}.
Figure \ref{fig:dist} shows the effect of this evolution.

\subsection{Emissivities and Absorptivities}

In the rest frame of the fluid, the synchrotron emissivity and absorptivity are defined with respect to the particle distribution function, $N_e(\gamma_e)$, and the single particle power spectrum, $P_e'$: 
\begin{equation}
j_{\nu'} = \frac{1}{4 \pi} \int d \gamma_e N(\gamma_e) P_e'(\gamma_e)
\end{equation}
and 
\begin{equation}
\alpha_{\nu'} = \frac{1}{8 \pi m_e \nu'^2} \int d\gamma_e \frac{N(\gamma_e)}{\gamma_e^2} \frac{d}{d\gamma_e}\left(\gamma_e^2 P_e'(\gamma_e)\right).
\end{equation}
For $P_e'$, we use the single particle power spectrum averaged over pitch angles:
\begin{equation}
P_e' = \frac{\sqrt{3}e^3 B}{m_e c^2}\int\limits^{\pi/2}_{0} d \alpha \sin(\alpha)^2 F\left[X/\sin(\alpha)\right]
\end{equation}
with 
\begin{equation}
X \equiv \frac{\nu'}{\nu_{synch}'(\alpha = \pi/2)} = \frac{ 4 \pi \nu' m_e c}{3 e B \gamma_e ^2}
\end{equation}
and 
\begin{equation}
F(y) = y \int \limits_y^{\infty} K_{5/3}(z)dz.
\end{equation}


\subsection{Radiative Transfer}

To determine the observed spectrum, 
we numerically solve the radiative transfer equation in the lab frame: 
\begin{equation}
\frac{d I_\nu}{d s } = j_\nu - I_\nu \alpha_\nu,
\end{equation}
which has been written in terms of the lab frame frequency, $\nu$, and the line of sight, $s$. We can rewrite the transfer equation in terms of fluid rest frame quantities $\nu'$, $j_{\nu'}$, and $\alpha_{\nu'}$ using the invariance of $j_\nu/\nu^2$ and $\alpha_\nu \nu$ : 
\begin{equation}
\frac{d I_\nu}{d s } = \left(\frac{\nu(1+z)}{\nu'}\right)^2 j_{\nu'}  -  \left(\frac{\nu'}{\nu(1+z)}\right) I_\nu \alpha_{\nu'},
\label{eq:rad_trans}
\end{equation}
where $\nu'/\nu = \gamma(1- \beta \mu) \approx \left[(1-\mu) \gamma +  \mu /(2\gamma)\right] (1+z)$ in terms of $\mu \equiv \cos(\theta)$, where $\theta$ is the polar angle, and $z$ is the redshift. Here we approximate the distance along the line of sight as $s = r \mu $, so that the relationship between $r$ and $t$ along a line of sight is given by 
\begin{equation}
t_{los} = \frac{ t_{obs}}{1+z} + \frac{r_{los} \mu(r_{los}, \theta_{obs})}{c},
\label{eq:los_def}
\end{equation}
for an observing time $t_{obs}$, defined such that $t_{obs} = 0$ corresponds to emission from photons at the origin of the blast wave at $t=0$. Along a line of sight, $\mu$ is function both of radius and the viewing angle $\theta_{obs}$, defined such that $\theta_{obs} = 0$ corresponds to the line of sight at the equator. In terms of $x \equiv \sin(\theta_{obs}) d_L/R_{\perp,max}$, where $d_L$ is the luminosity distance to the source and $R_{\perp,max}$ is the maximum value of $r \sin(\theta)$ over all lines of sight, we have
\begin{equation}
\mu = \sqrt{1 - x^2 \left(\frac{R_{\perp,max}}{r}\right)^2},
\end{equation}
For a spherical blast wave (as we use here) equation \eqref{eq:los_def} defines an ovoidal shape for the surface of equal arrival times of photons emitted by the flow. For a detailed and useful discussion of the properties of this surface, we refer the reader to \citet{GPS1999} (where it is referred to as the ``egg'').  

After solving equation \eqref{eq:rad_trans}, we obtain the flux by integrating the specific intensity over solid angles:
\begin{equation}
F_\nu = \int I_\nu d\Omega  = 2 \pi (1+z)\left(\frac{R_{\perp,max}}{d_L}\right)^2\int\limits_0^1 I_\nu xdx.
\end{equation}

\section{Results}

\begin{figure}
\centering
\includegraphics[width=0.45\textwidth]{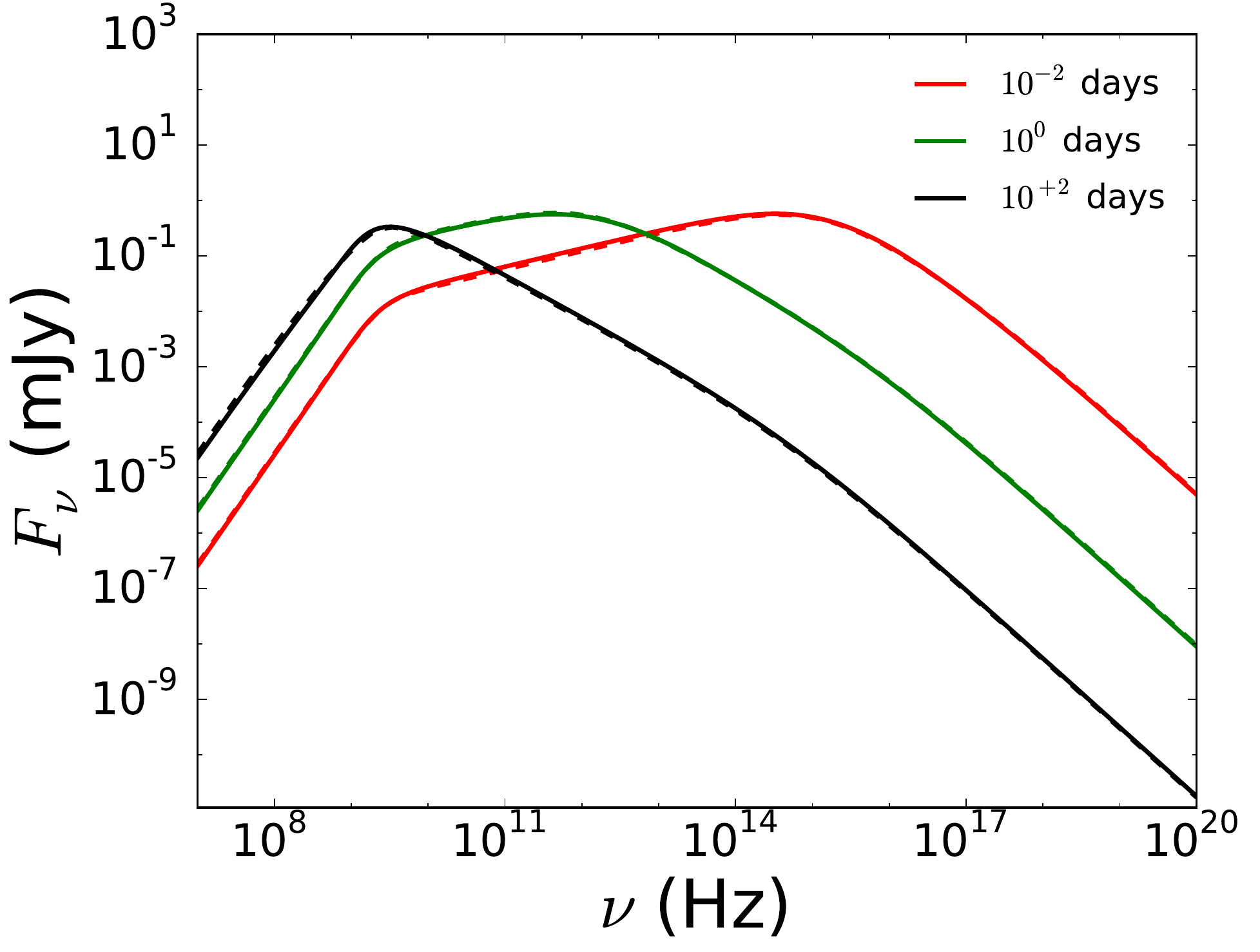}
\caption{Spectrum of the nonthermal particles only at several different times for our fiducial parameters, computed with our code (solid lines) compared to the fitting formulae of GS02 (dashed lines). The differences are predominantly  $ \lesssim 10\%$, with the largest difference ($\sim 20\%$) occurring in the self-absorbed segment in the spectrum at $10^2$ days. At this time, the GS02 prescription is not well defined and so we instead compare our calculation to the modified prescription of \citet{lbt+14}: a phenomenological weighted average of two GS02 spectra not expected to precisely match the full calculation. 
}
\label{fig:gs_comp}
\end{figure}

We select fiducial values of $E' = 10^{52}$\,erg, $n_{ext}' = 1\,{\rm cm}^{-3}$, $\epsilon_e'=0.1$, $\epsilon_B'=0.01$, and $p=2.5$ for our analysis. Since synchrotron radiation from nonthermal electrons with the parameter sets [$E'$,$n_{ext}'$,$\epsilon_e'$,$\epsilon_B'$; $\fNT = 1$] and [$E'/\fNT$,$n_{ext}'/\fNT$,$\fNT\epsilon_e'$, $\fNT\epsilon_B'$; $\fNT < 1$] is identical, we fix the properties of the nonthermal particles by scaling the parameters with \fNT\ in this way to study the effect of the thermal population \citep{Eichler2005}.  For the cold electron model, we fix $\eta_e = 10^{-3}$, corresponding to $\Theta_{e,0} = 10^{-2} \gamma_{\rm min,0}$ (using equations \ref{eq:gamma_min} and \ref{eq:theta_eta}).

We note that for these parameters, the emission at $\sim 10^{2}$ days is dominated by photons emitted when $\Gamma$ was $\sim $ a few. A break in the light curves due to the finite opening angle of the outflow ($\theta_0$) is expected when $\Gamma \theta_0 \sim 1$ \citep{sph99}. We neglect this effect by considering a spherical blast wave, in order to particularly isolate the effect of adding thermal electrons to the framework of GS02.


We test our calculations for the case of $\fNT = 1$ (no thermal electrons) against the basic features of the synchrotron model comprising power law spectral segments connected at spectral break frequencies: the synchrotron self absorption frequency, \nua, the characteristic frequency, \numax, and the cooling frequency, \nuc. Our results reproduce the expected spectral segments and are consistent with the fitting formulae of GS02 (with modifications as described by \citealt{lbt+14}) within $\approx20\%$ at all frequencies (Figure \ref{fig:gs_comp}). 

\begin{figure*}
\includegraphics[width=0.32\textwidth]{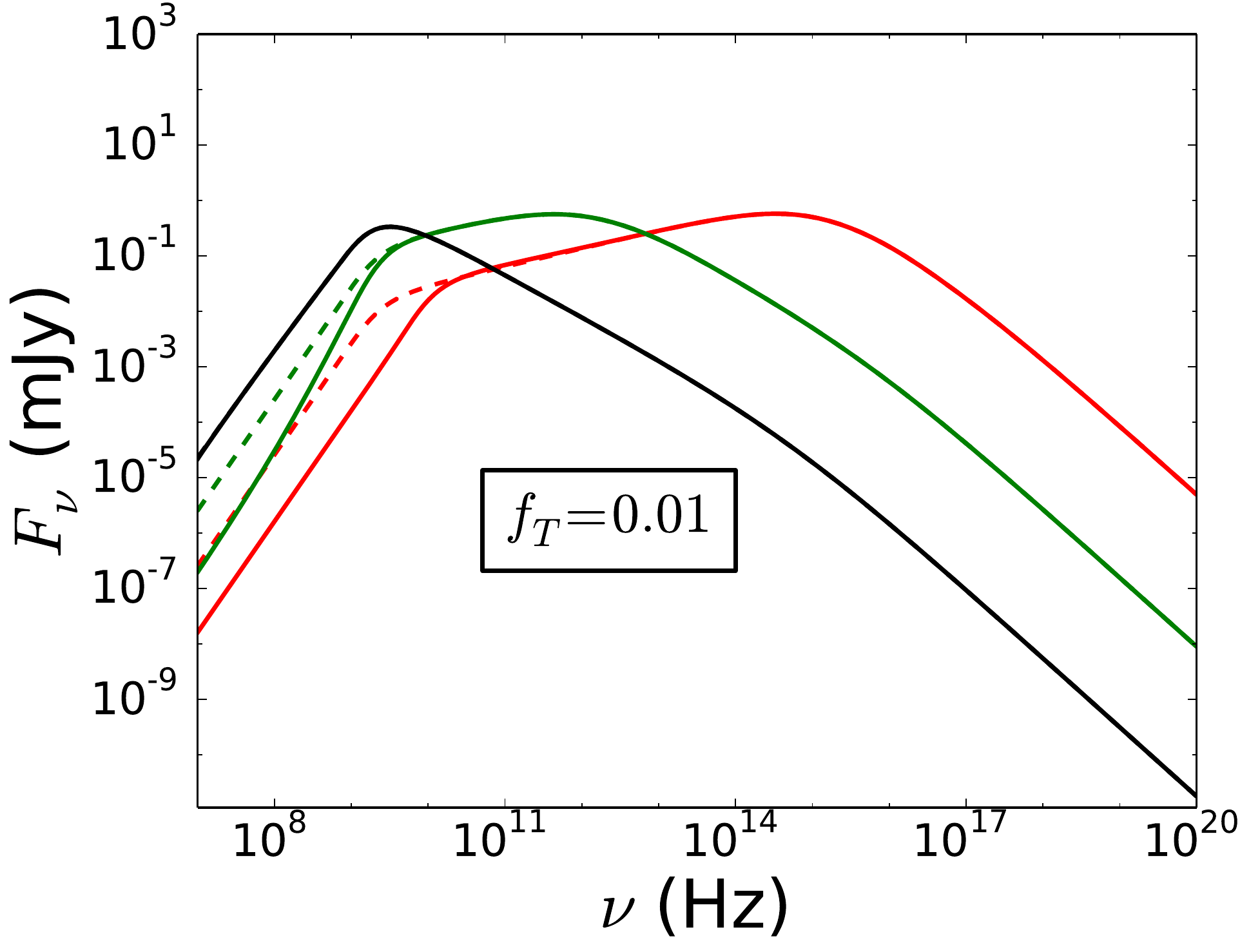}
\includegraphics[width=0.32\textwidth]{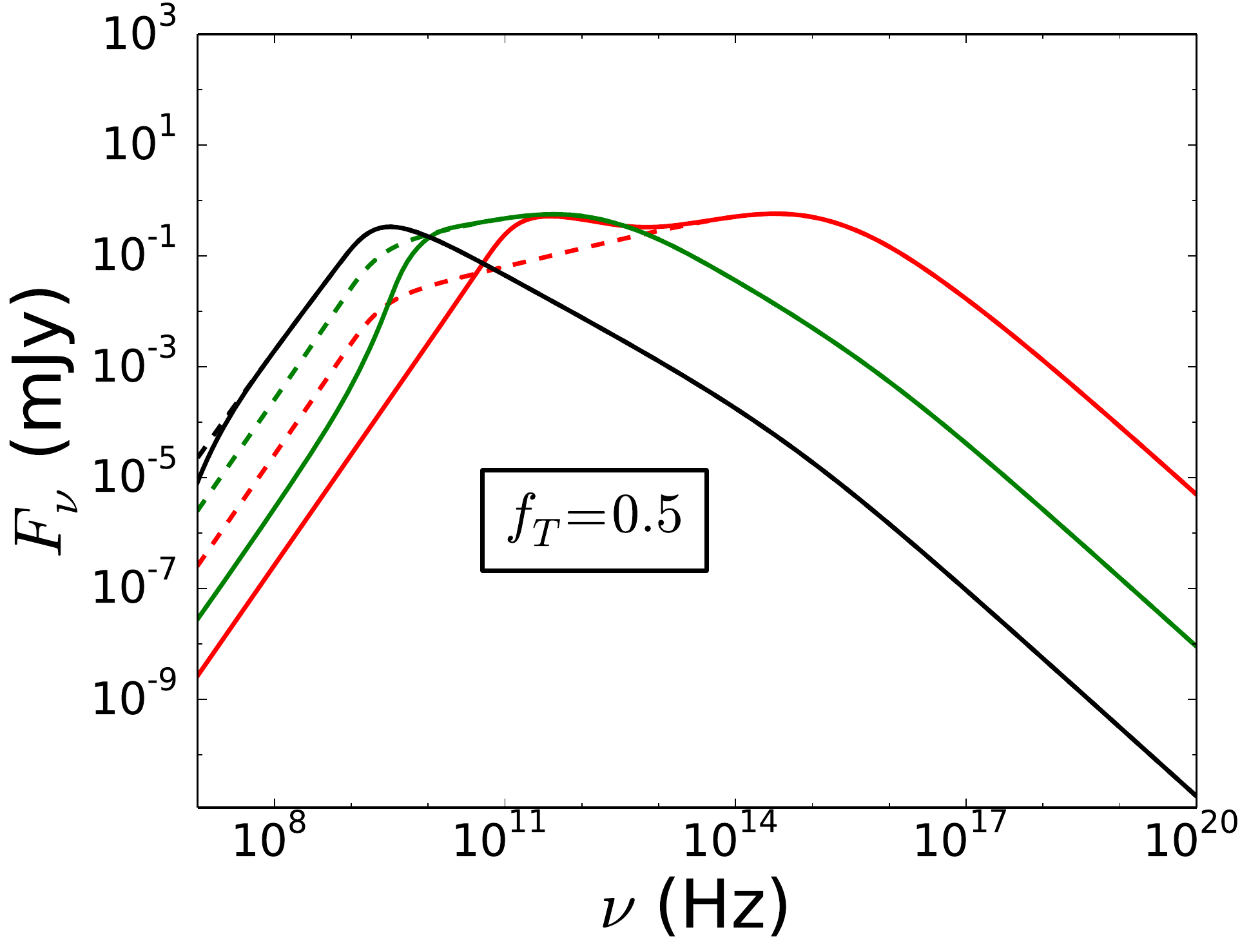}
\includegraphics[width=0.32\textwidth]{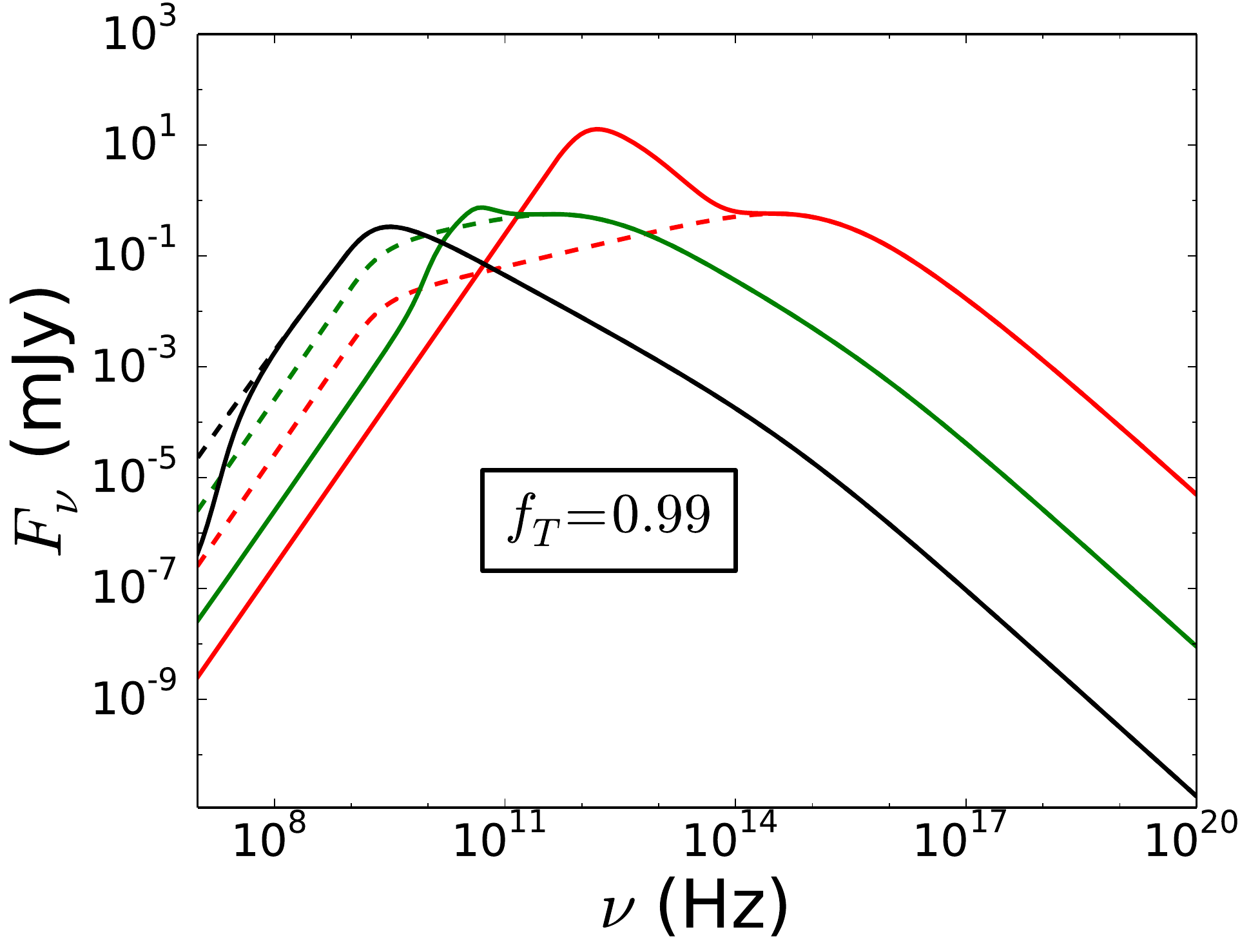}\\
\includegraphics[width=0.32\textwidth]{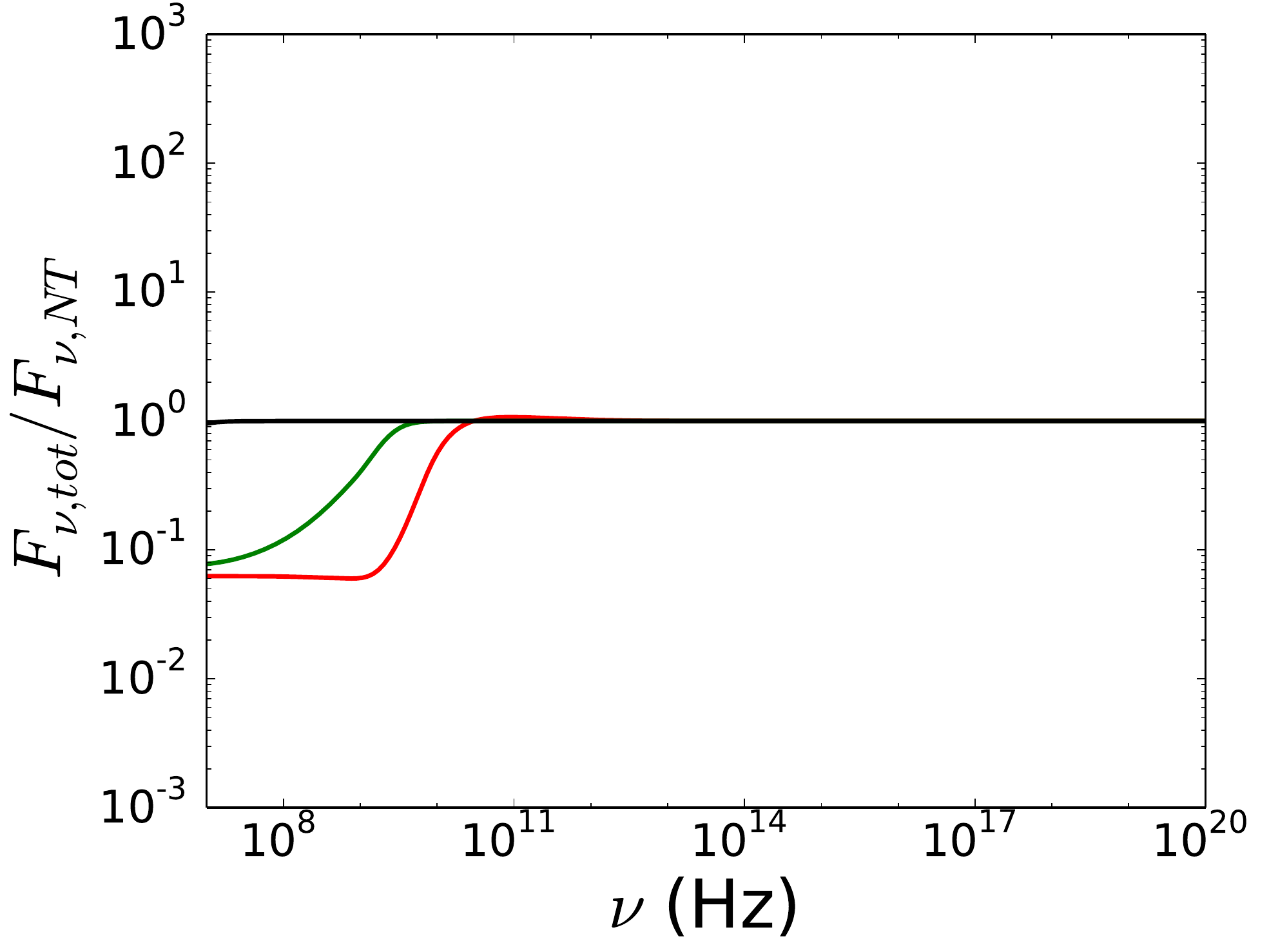}
\includegraphics[width=0.32\textwidth]{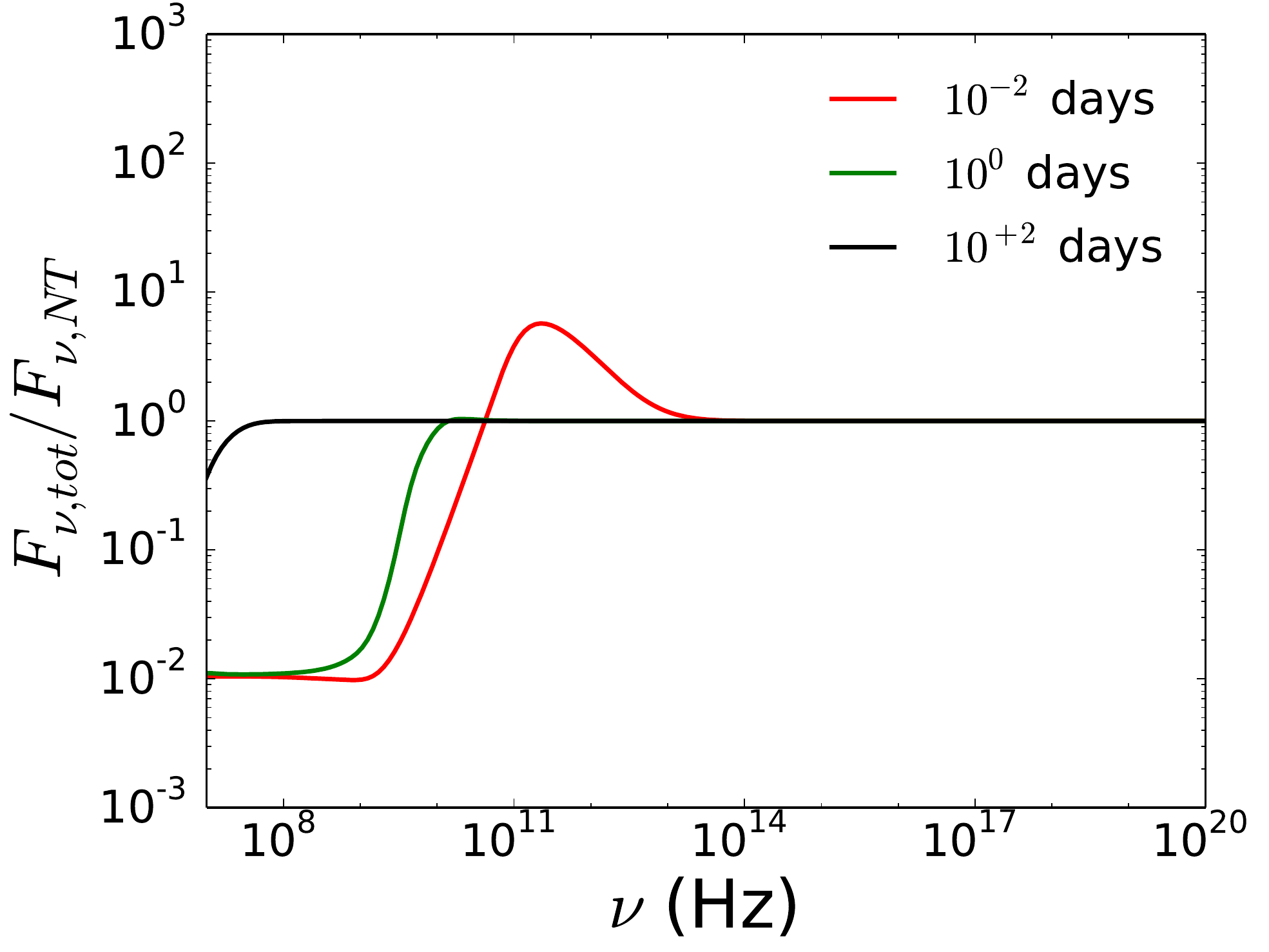}
\includegraphics[width=0.32\textwidth]{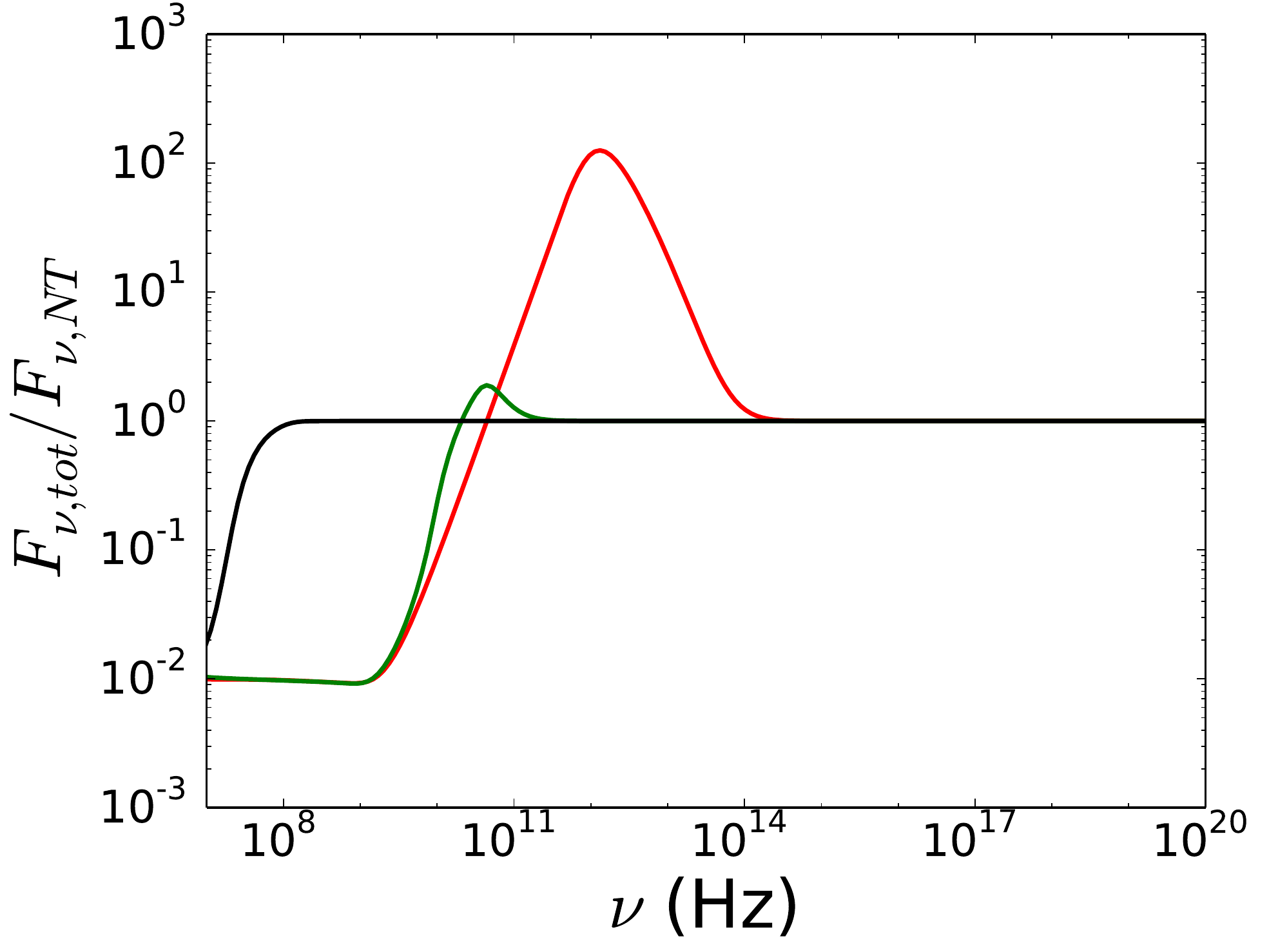}
\caption{Top: Spectral energy distributions for the nonthermal particles (dashed) and all particles (solid) in the cold electron model (\S \ref{sec:cold}) at three  observation times for different thermal electron fractions.
The electron temperature is a fixed fraction ($\eta_e = 10^{3}$) of the gas temperature, so that $\Theta_{e,0} = 10^{-2} \gamma_{\rm min,0}$. 
Bottom: Ratio between the total and nonthermal spectra. The presence of thermal particles significantly suppresses the optically thick emission, increases \nua, and generates an additional emission component above \nua. }
\label{fig:dis_spec}
\end{figure*}

In the cold electron model, the introduction of even a small fraction of non-accelerated particles ($\fT=0.01$) significantly increases the optical depth to synchrotron self-absorption, creating a deficit in the observed radiation compared to the case with no thermal electrons (Figure \ref{fig:dis_spec}). This flux-suppression ranges from $\sim 10^{-1}$ at $\fT=0.01$ to $\sim 10^{-2}$ at $\fT = 0.5$ and $0.99$,
and the absorption is dominated by thermal particles.
For $\fT = 0.01$, the thermal particles are primarily absorbing emission from nonthermal particles, while for $\fT = 0.5$ and $0.99$, the thermal particles are primarily absorbing their own radiation.  This increased optical depth also leads to a higher effective self-absorption frequency when $\nuaT>\nuaNT$, where $\nuaNT$ (\nuaT) is the self-absorption frequency considering only absorption by nonthermal (thermal) particles. While \nua\ appears to be well approximated by $\nua \approx \max (\nuaNT,\nuaT)$, it is in general a nonlinear function of \nuaT\ and \nuaNT, and we leave the precise determination of an expression for \nuaNT\ and \nua\ to future work.  
Furthermore, when $\numaxT<\nu<\nua$, the rapid decline of the thermal absorptivity above $\numaxT$ leads to a steepening of the
spectrum over the transition from thermal- to nonthermal-dominated absorption.

Additionally, for larger values of $\fT$ ($\gtrsim 0.5$), we find excess flux between \nuaNT\ and \numaxNT\ from the thermal population, characterized by a $\nu^{2}$ self-absorbed segment breaking into $\nu^{1/3}$ at \nua, followed by a rapid decline above the peak at \numaxT. The latter is related to \numaxNT\ by
\begin{equation}
\numaxT \approx \left(\frac{\eta_e \fNT }{3\epsilon_e }\right)^2 \left(\frac{p-2}{p-1}\right)^2 \numaxNT,
\end{equation}
which holds for all values of \fT. This expression is derived from the relationship between $\Theta_{e,0}$ and $\gamma_{min,0}$ (see equations \ref{eq:gamma_min} and \ref{eq:theta_eta}), and the expected scaling of $\numax' \propto \gamma_e^2 B$.   
Whereas the excess emission fades with time, the increased optical depth to self-absorption persists.

\begin{figure*}
\includegraphics[width=0.32\textwidth]{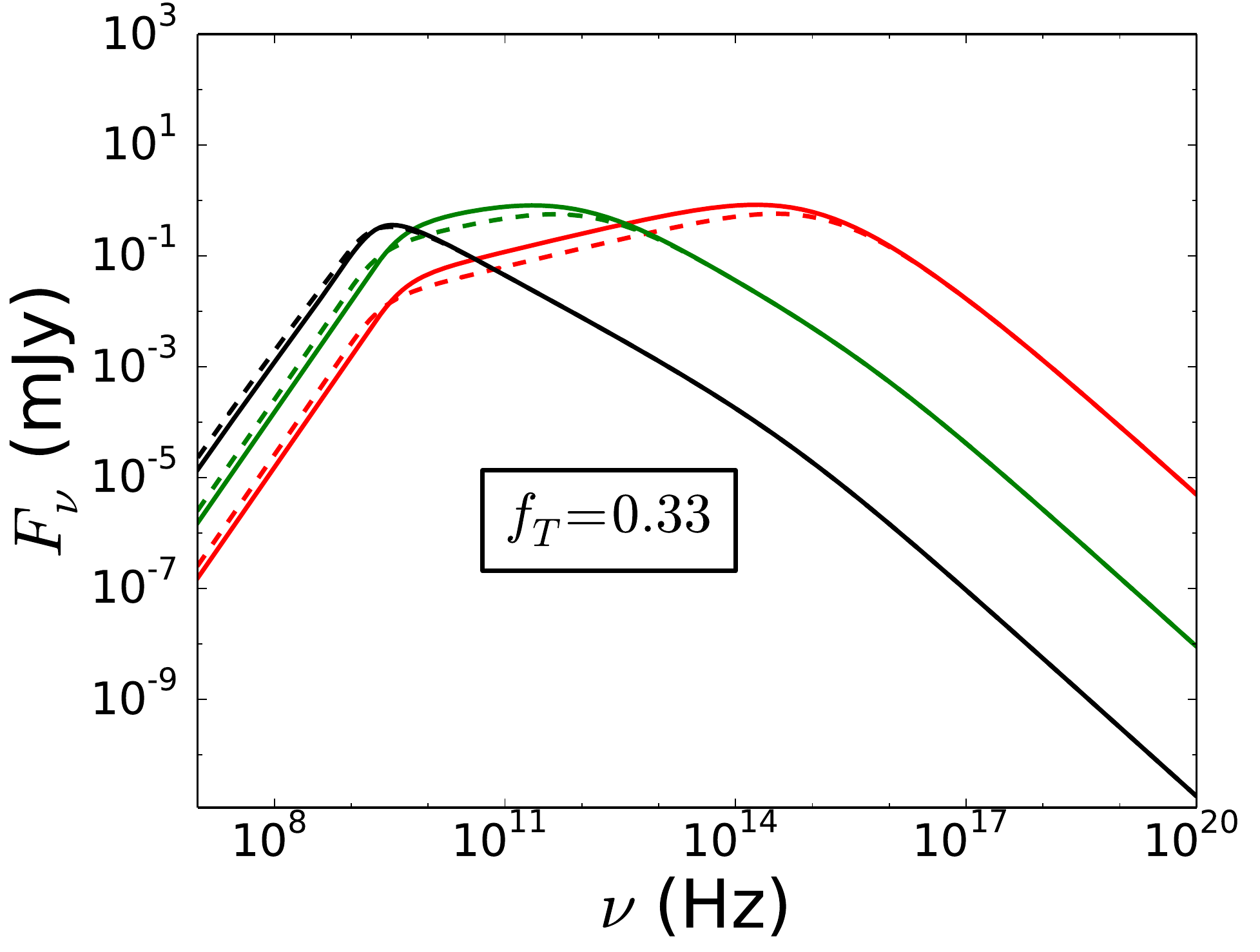}
\includegraphics[width=0.32\textwidth]{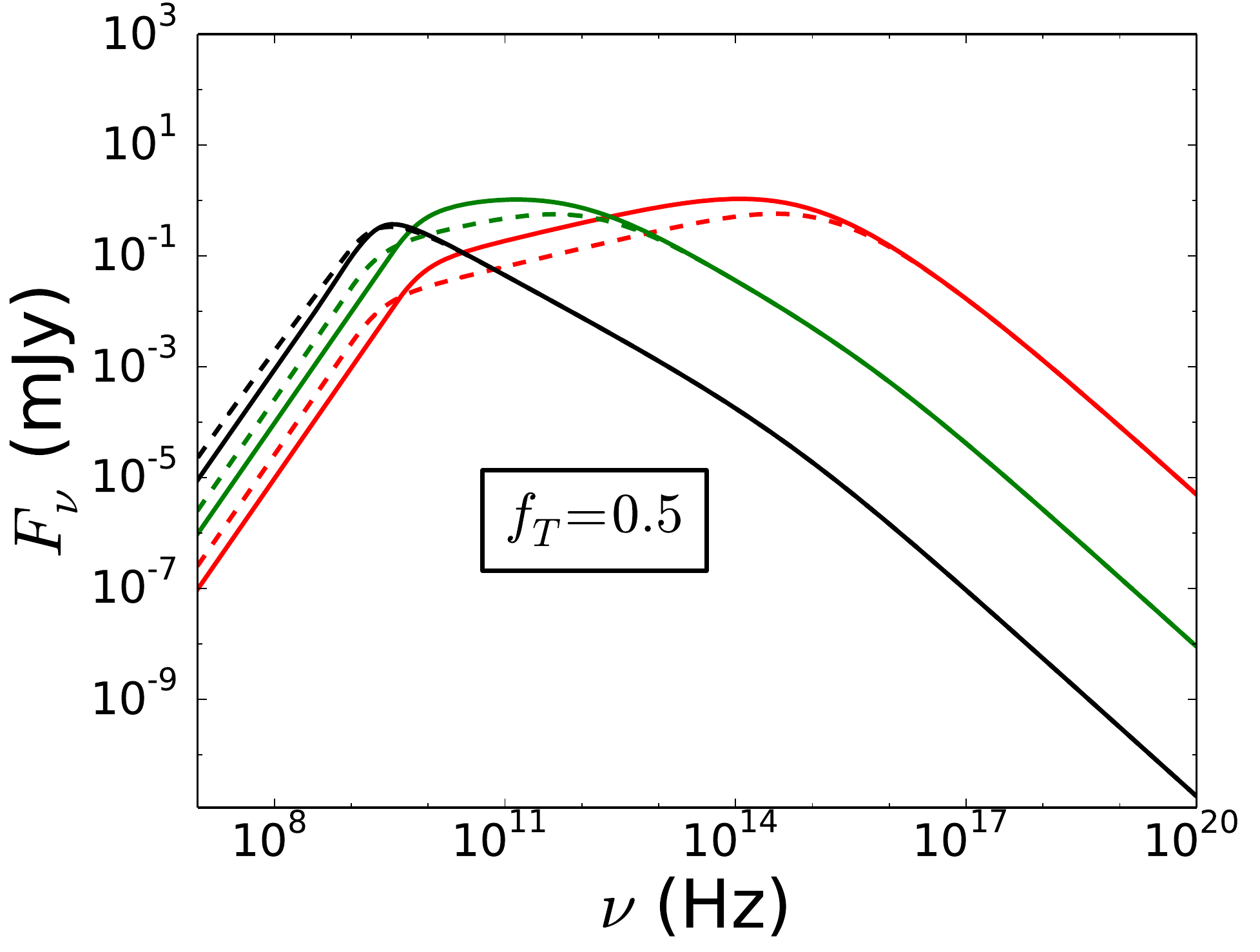}
\includegraphics[width=0.32\textwidth]{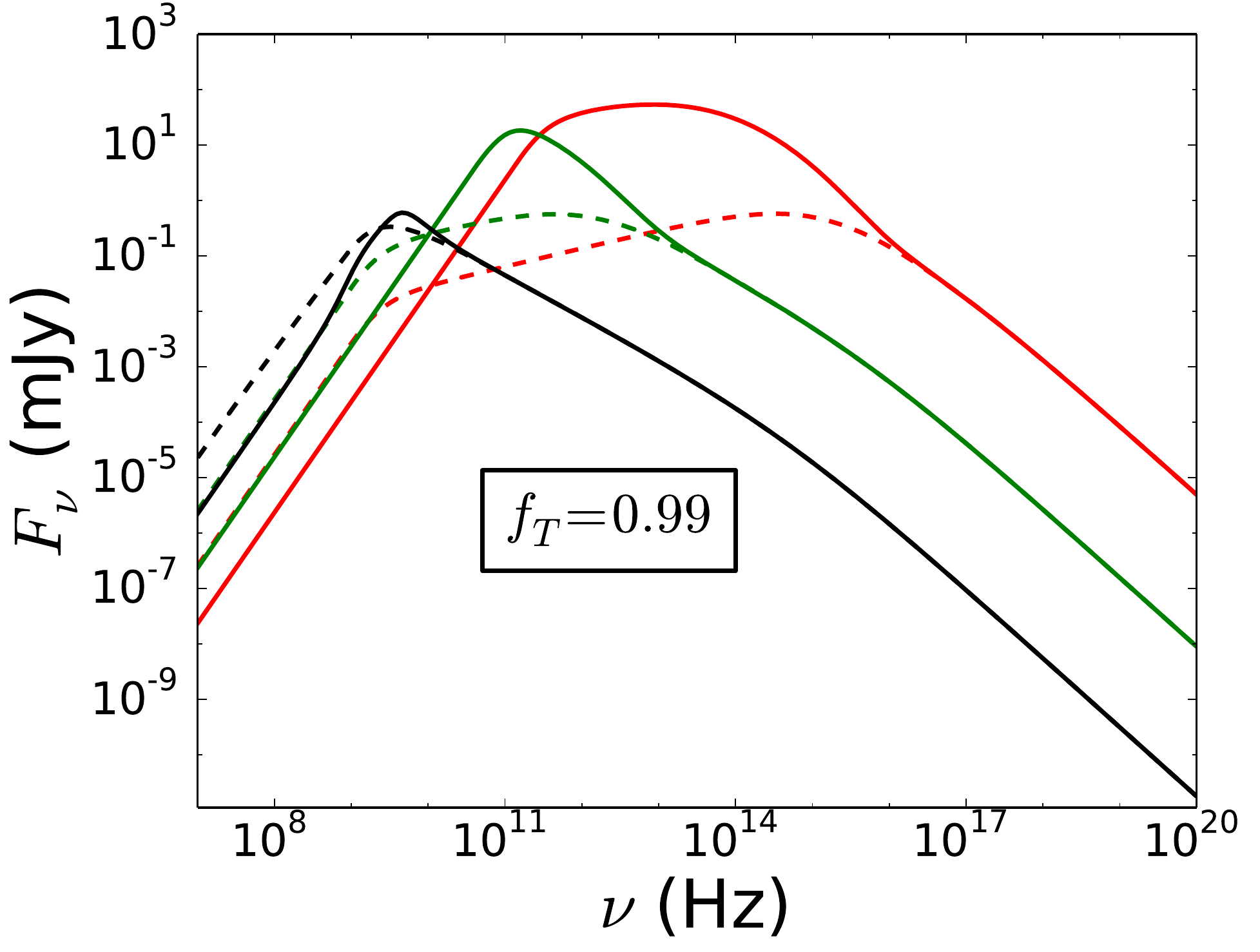}\\
\includegraphics[width=0.32\textwidth]{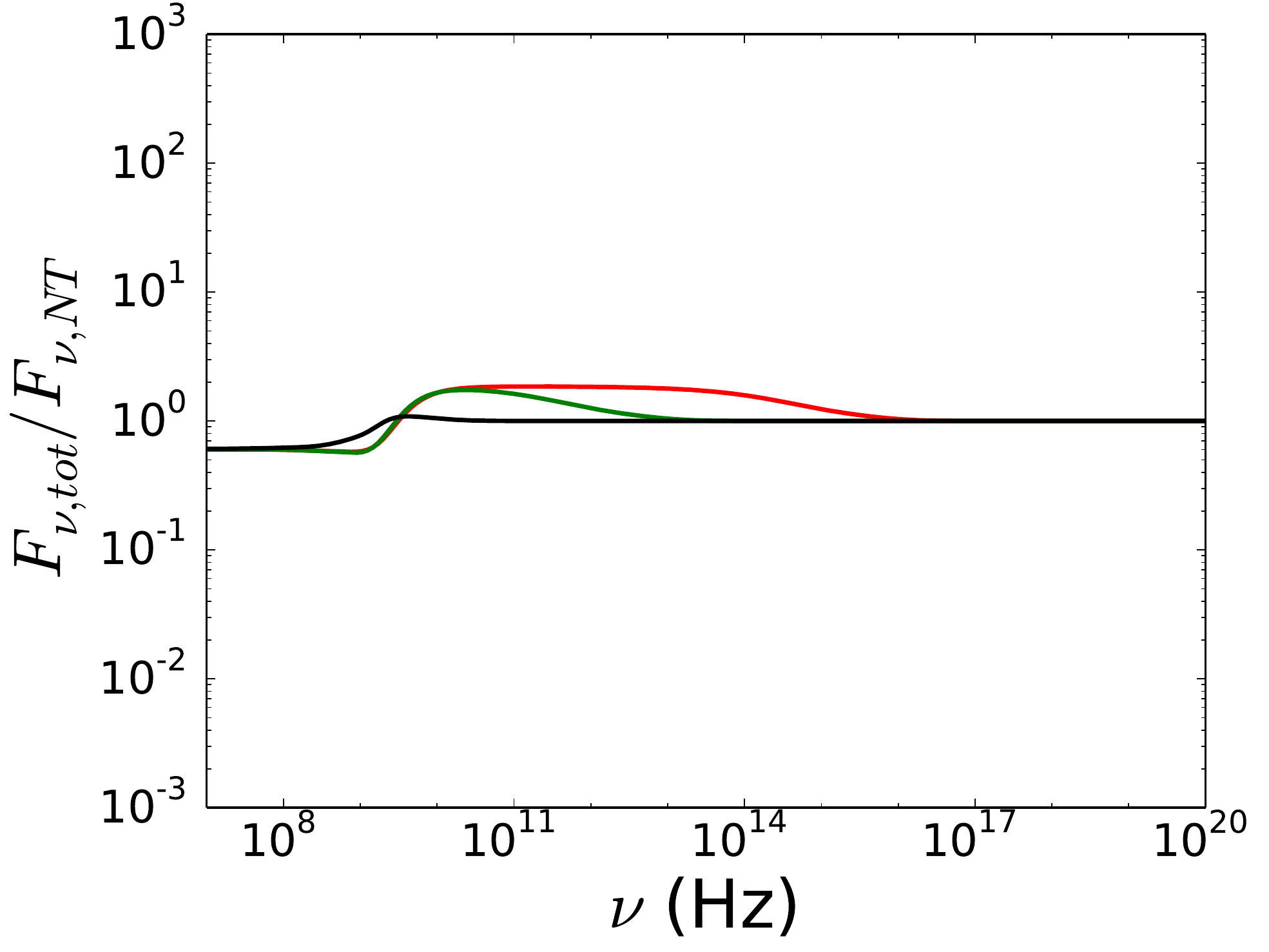}
\includegraphics[width=0.32\textwidth]{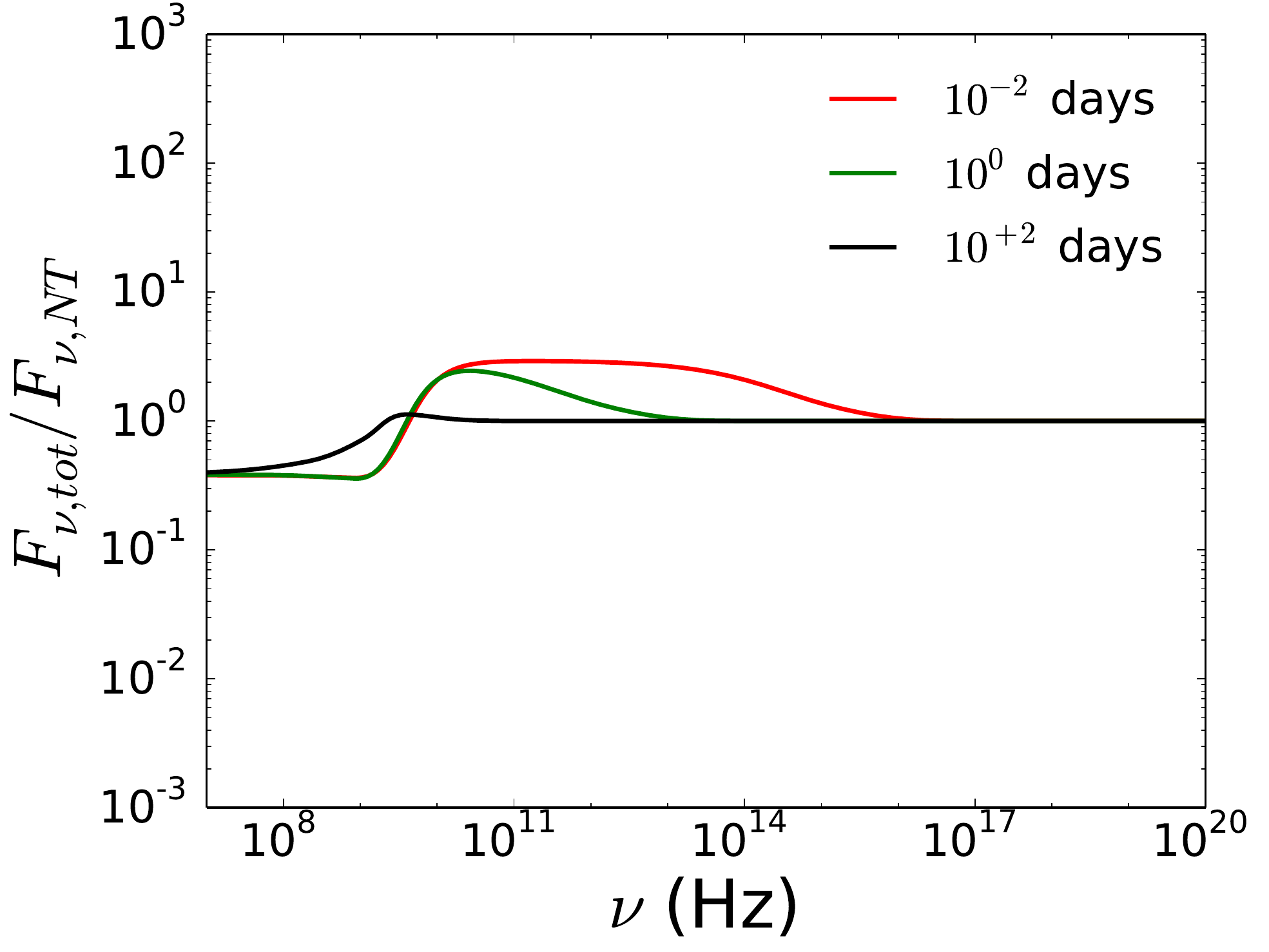}
\includegraphics[width=0.32\textwidth]{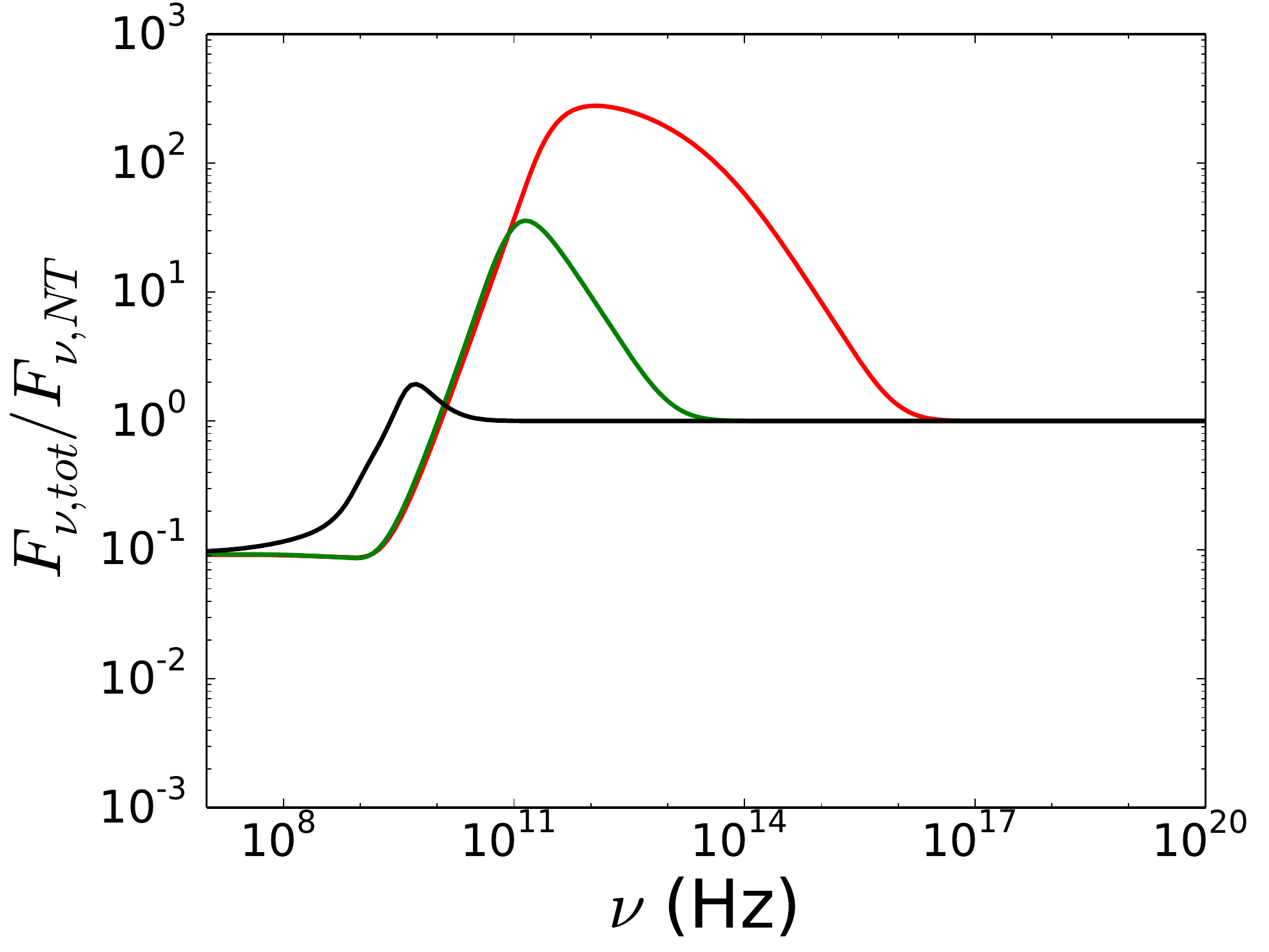}
\caption{
Same as Figure \ref{fig:dis_spec} but for the hot electron model (\S \ref{sec:hot}). Note that below $\fT\approx 0.33$ the thermal and nonthermal distributions can no longer be continuously matched at $\gamma_{\rm min,0}$ for this model. 
Compared to the cold electron model, the thermal particles in the hot electron model radiate up to higher frequencies and provide less reduction in optically thick emission due to the higher temperatures.
}
\label{fig:cont_spec}
\end{figure*}

We find that these same effects are seen in the hot electron model, albeit to a varying extent.  The observable effects of $\fT>0$ are smaller below $\nua$ compared to the cold electron model, whereas the excess emission (between \nuaNT\ and \numaxNT) extends to higher frequencies (Figure \ref{fig:cont_spec}). Both effects can be traced to the higher effective temperature of the thermal electrons, which results in an increased effective blackbody temperature and higher \numaxT.  Note that requiring continuity of the distribution function results in the thermal Maxwellian being only a small correction to the nonthermal distribution for $\fT \lesssim 0.8$ (e.g., $\fT = 0.5$ in Figure \ref{fig:dist}), which may indicate that higher values of \fT\ are more realistic.

\begin{figure*}
\includegraphics[width=0.49\textwidth]{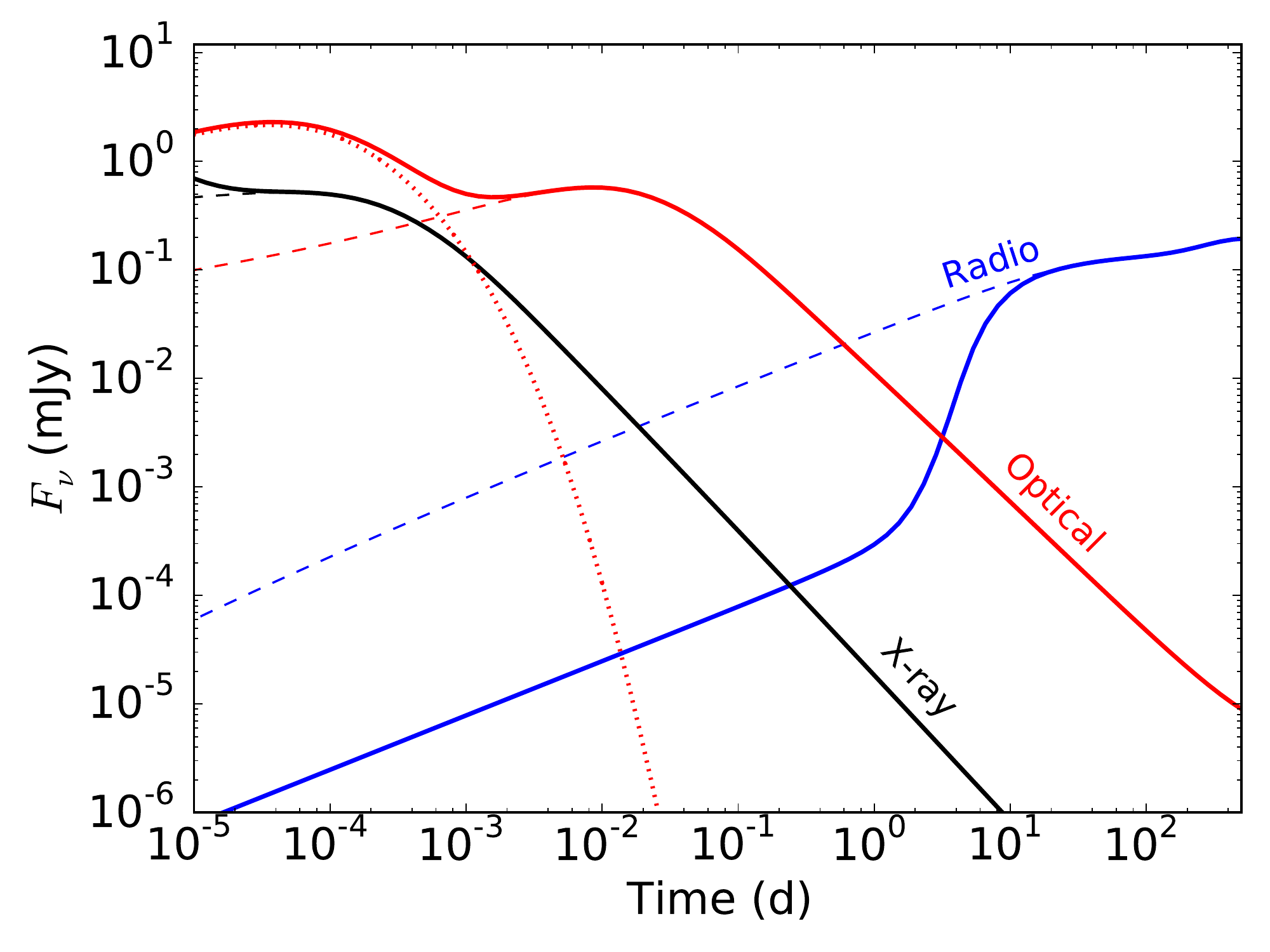}
\includegraphics[width=0.49\textwidth]{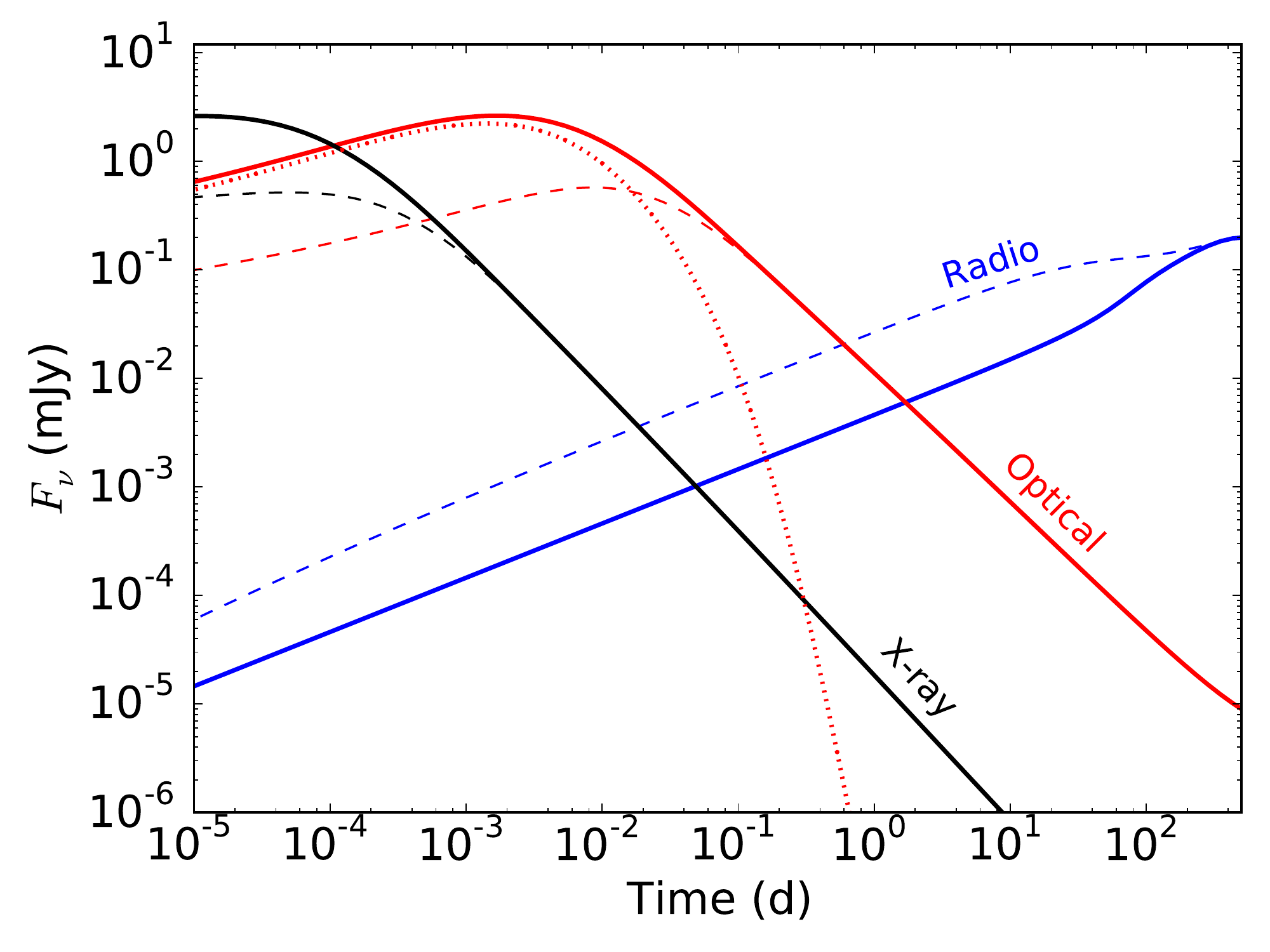}
\caption{1\,keV X-ray band (black), optical $i^{\prime}$-band (red), and radio \textit{L}-band (blue) light curves of synchrotron radiation ($2 \times 10^{17}$ Hz, $4 \times 10^{14} $ Hz, and $10^9$ Hz, respectively) with our fiducial set of parameters for the nonthermal particles (dashed), thermal particles (dotted; only shown for the optical), and the total (solid), assuming $\fT=0.8$ for the cold electron model (left) and hot electron model (right).  For both models, the inclusion of emission and absorption from thermal electrons leads to an excess of optical and X-ray emission at early times, lasting longer for the hot electron model.  There is also a persistent reduction in radio emission, stronger for the cold electron model, which lasts until late times when $\numaxT$ drops below the the radio \textit{L}-band. The difference in magnitude of these effects for the hot and cold electron models is as expected from the difference in the assumed post-shock electron temperature.  }
\label{fig:lc}
\end{figure*}

We show sample light curves in the X-ray, optical, and radio for both models with $\fT = 0.8$ in Figure \ref{fig:lc}.  Synchrotron radiation from thermal electrons results in excess emission in the optical at early times ($\lesssim 1$\,d), and the effect lasts longer for the hot electron model. The additional absorption from the thermal particles causes a rapid rise in the radio light curve when \numaxT\ passes through and the optical depth of the thermal electrons rapidly declines.  This effect is more pronounced for the cold electron model and happens at an earlier time. 

To quantify the impact of ignoring thermal electrons in parameter estimation during afterglow modeling, we perform two sample Monte Carlo analyses of realistic X-ray, UV/optical, and radio data sets generated from the $\fT=0.8$ cold and hot electron models (including noise, instrumental sensitivity limits, typical cadences, and scintillation effects), which we subsequently fit with the GS02 formalism assuming $\fNT=1$. The resulting parameter estimates are expected to differ from their input values due to both the degeneracy arising from $\fNT<1$ \citep{Eichler2005}, and the spectral differences caused by the additional thermal distribution of electrons. In order to distinguish among the two, we compare the fitted parameters both to the input values (``True'') and the equivalent (``expected'') values for $\fNT=1$ in Table \ref{tab:params}.  Comparison of the fitted values to the ``True'' values includes both sources of error, while comparison to the ``expected'' values includes only the errors incurred by the spectral differences.
For both models, we find that ignoring the contribution of thermal particles causes errors in the determination of the physical parameters by 10\%-500\% relative to the true values and 10\%-400\%  relative to the expected values. 
For both models, the errors relative to the expected values are highest for the density due to the increased optical depth from the thermal electrons; however, this error partially cancels the effect of the degeneracy due to $\fNT<1$, bringing $n_0$ closer to the true value.

Additionally, in the hot electron model we find significantly larger errors than in the cold electron model in the fits for $\epsilon_e$ and $E$ compared to the expected values (factors of $\approx$ 3.5 and 33, respectively), caused by the higher temperature electrons having an effect on more of the higher frequency portion of the spectrum. The error in $E$ relative to the true value behaves similarly to the error in $n_0$ in that it is partially cancelled by the degeneracy uncertainty.  The error in $\epsilon_e$, on the other hand, behaves in the opposite way, and in fact increases relative to the true value when the degeneracy uncertainty is included.  We conclude that both the parameter degeneracy for $\fNT<1$ and the spectral differences caused by $\fT>0$ can lead to significant sources of error that interact in complex ways.  In some cases, the errors can partially cancel out, but in others they can compound; the interplay between the two will likely depend upon the burst properties, as well as the time and frequency sampling of of afterglow data. Therefore, in order to properly constrain the parameters of GRB afterglows with $\fNT<1$, both effects need to be taken into account using a model for the thermal electron emission and absorption. 


\begin{deluxetable}{lcccc}
 \tabletypesize{\footnotesize}
 \tablecolumns{5}
 \tablecaption{MCMC Parameter fits  }
 \tablehead{   
           \colhead{Parameter} &
           \colhead{True} &
           \colhead{Expected} &
           \colhead{Cold} &
           \colhead{Hot}
   }
 \startdata       
   $p$            & 2.5 & 2.5 & $2.517\pm0.004$             & $2.486 \pm 0.007$ \\[2pt]
   $\epsilon_{e}$ & $2\times10^{-2}$ & 0.1 & $0.119\pm0.004$ & $(3.4\pm0.5)\times10^{-2}$ \\ [2pt]
   $\epsilon_{B}$ & $2\times10^{-3}$ & 0.01 
                  & $(3.7\pm0.4)\times10^{-3}$          & $(2.1\pm0.6)\times10^{-3}$ \\ [2pt]
   $n_0$          & 5.0  & 1.0  & $4.1^{+0.8}_{-0.4}$     & $5.4^{+5.2}_{-2.4}$ \\[2pt]
   $E_{\rm 52}$\tablenotemark{a} & 5.0 & 1.0 & $0.89\pm0.03$         & $4.6\pm0.4$ \\[2pt]
   $f_{\rm NT}$   & 0.2 & 1.0 & 1.0\tablenotemark{b}   & 1.0\tablenotemark{b}
 \enddata 
  \tablenotetext{a}{$E_{52}$ is $E$ in units of $10^{52}$ ergs.}
  \tablenotetext{b}{\fNT\ is held fixed during the analysis.}
 \tablecomments{Parameter fits to mock observations generated from our cold and hot electron models.  The parameter estimation is done by fitting the mock observations to the GS02 model which includes nonthermal particles only. ``Expected'' values are the \fNT=1 equivalent values of the ``True'' input parameters.  Neglecting the contribution of thermal emission can lead to errors in parameter estimation beyond even the known degeneracy in the nonthermal spectrum for $\fNT<1$.  }

\label{tab:params}
\end{deluxetable}

\section{Discussion and Conclusions}
We have expanded the standard GRB afterglow model to include the additional contribution of a thermal distribution of electrons. We have shown that this additional population of electrons generally has two effects on the spectrum. The first is an excess of flux that occurs near the peak synchrotron frequency of the thermal electrons. This additional component fades with time as the emission shifts to lower frequencies, consistent with the results of \citet{gs09} and \citet{webn17}. Secondly, the optically thick low frequency radio emission is reduced from what one would expect for a purely nonthermal electron distribution. The strength of both effects depend on the post-shock temperature of the thermal population relative to that of the nonthermal population. The detection of these features in the spectra of GRB afterglows could be an indication of inefficient electron acceleration. 


We have shown that neglecting the contribution of thermal electrons in modelling observational GRB afterglow sources introduces errors in parameter estimation in addition to the known degeneracy in parameters in nonthermal electron-only spectra when $\fNT<1$. Future detailed explorations of the effect of the acceleration efficiency and the relative temperature of the thermal and nonthermal components could be used to generate fitting functions, thereby facilitating multi-wavelength modeling of GRB afterglows. This model could then be used to break the parameter degeneracy by obtaining a direct measurement of the fraction of accelerated electrons, thus informing studies of electron acceleration in collisionless relativistic shocks.

Whereas our study is focused on the two extreme cases of either strong or weak thermal shock heating (i.e., the hot and cold electron models, respectively), it is likely that shocks in GRB afterglows generally fall somewhere in between these two limiting cases. Our analysis quantitatively demonstrates the possible range of effects caused by a non-negligible fraction of non-accelerated electrons in afterglow shocks in order to better interpret observations and lay the groundwork for more detailed future modeling.


\section*{Acknowledgments}
We thank Jonathan Granot, Eliot Quataert, Edo Berger, and Re'em Sari for useful discussions, and the anonymous referee for their suggestions to improve the manuscript.  SMR is supported in part by the NASA Earth and Space Science Fellowship.  TL is a Jansky Fellow of the National Radio Astronomy Observatory. This work was made possible by computing time granted by UCB on the Savio cluster.



\bibliographystyle{mn2efix}

\end{document}